\documentclass[prd,preprint,superscriptaddress,amsmath,amssymb,nofootinbib]{revtex4}
\usepackage{graphicx}% Include figure files
\usepackage{dcolumn}% Align table columns on decimal point
\usepackage{bm}% bold math
\usepackage{amssymb}
\usepackage{amsmath}
\usepackage{epsfig}    
\usepackage{color}
\usepackage{slashed}
\usepackage{hhline}
\def\be{\begin{equation}}
\def\ee{\end{equation}}
\newcommand{\bea}{\begin{eqnarray}}
\newcommand{\eea}{\end{eqnarray}}
\newcommand{\nn}{\nonumber}

\numberwithin{equation}{section}

\begin{document}

{\begin{flushright}{APCTP Pre2021-022}
\end{flushright}}

\title{Zee-Babu model in  modular $A_4$ symmetry}

\author{Hiroshi Okada}
\email{hiroshi.okada@apctp.org}
\affiliation{Asia Pacific Center for Theoretical Physics, Pohang 37673, Republic of Korea}
\affiliation{Department of Physics, Pohang University of Science and Technology, Pohang 37673, Republic of Korea}

\author{Yong-hui Qi}
\email{yonghui.qi@apctp.org}
\affiliation{Asia Pacific Center for Theoretical Physics, Pohang 37673, Republic of Korea}

\pacs{}
\date{\today}

\begin{abstract}
We study a Zee-Babu model in a modular $A_4$ flavor symmetry, in which we search for several predictions such as phases, sum of neutrino masses, and neutrinoless double beta decay,  satisfying neutrino oscillation data in a minimum framework of the charge assignments of modular weight. We perform $\Delta \chi^2$ analysis to get our results and find $\tau$ is localized nearby at one of the fixed points of $i\times \infty$ for both of normal and inverted mass hierarchies.  
\end{abstract}

\maketitle

\section{Introduction}
Zee-Babu model is one of the attractive scenarios to get active neutrino matrix at two-level, since only one singly- and one doubly-charged bosons are introduced in the standard model(SM) without any additional symmetries~\cite{Zee-Babu}.
Also the neutrino mass matrix depends on structure of the charged-lepton mass matrix.
Thus, it has a lot of interesting phenomenologies that should be taken in account; e.g., lepton flavor universalities~\cite{Babu:2002uu}, lepton flavor violations~\cite{Babu:2002uu, AristizabalSierra:2006gb}, collider physics at Large Hadron Collider for new charged-bosons~\cite{Schmidt:2014zoa, Nebot:2007bc}.
The model  has a unique prediction for lepton sector that one of the lightest neutrino masses is always zero because the mass matrix consists of three by three anti-symmetric matrix whose rank is two.
However, there still exist too many free parameters to get further predictions such as CP phases and mixing patterns in the lepton sector.

A few years ago, an attractive flavor symmetry has been proposed by papers~\cite{Feruglio:2017spp,
deAdelhartToorop:2011re}, in which they have applied a non-Abelian discrete flavor
symmetry originated by a modular symmetry in order to find further predictions for quark and lepton sectors without introducing so many bosons.
%%%%%  定期分(start) %%%%%%% 
Along the line of this idea, a vast reference has recently appeared in the literature, {\it e.g.},  $A_4$~\cite{Feruglio:2017spp, Criado:2018thu, Kobayashi:2018scp, Okada:2018yrn, Nomura:2019jxj, Okada:2019uoy, deAnda:2018ecu, Novichkov:2018yse, Nomura:2019yft, Okada:2019mjf,Ding:2019zxk, Nomura:2019lnr,Kobayashi:2019xvz,Asaka:2019vev,Zhang:2019ngf, Gui-JunDing:2019wap,Kobayashi:2019gtp,Nomura:2019xsb, Wang:2019xbo,Okada:2020dmb,Okada:2020rjb, Behera:2020lpd, Behera:2020sfe, Nomura:2020opk, Nomura:2020cog, Asaka:2020tmo, Okada:2020ukr, Nagao:2020snm, Okada:2020brs, Yao:2020qyy, Chen:2021zty, Kashav:2021zir, Okada:2021qdf, deMedeirosVarzielas:2021pug, Nomura:2021yjb, Hutauruk:2020xtk, Ding:2021eva, Nagao:2021rio, king},
%%%
$S_3$ \cite{Kobayashi:2018vbk, Kobayashi:2018wkl, Kobayashi:2019rzp, Okada:2019xqk, Mishra:2020gxg, Du:2020ylx},
%%%
$S_4$ \cite{Penedo:2018nmg, Novichkov:2018ovf, Kobayashi:2019mna, King:2019vhv, Okada:2019lzv, Criado:2019tzk,
Wang:2019ovr, Zhao:2021jxg, King:2021fhl, Ding:2021zbg, Zhang:2021olk, gui-jun, Nomura:2021ewm},
%%%
$A_5$~\cite{Novichkov:2018nkm, Ding:2019xna,Criado:2019tzk}, double covering of $A_5$~\cite{Wang:2020lxk, Yao:2020zml, Wang:2021mkw, Behera:2021eut}, larger groups~\cite{Baur:2019kwi}, multiple modular symmetries~\cite{deMedeirosVarzielas:2019cyj}, and double covering of $A_4$~\cite{Liu:2019khw, Chen:2020udk, Li:2021buv}, $S_4$~\cite{Novichkov:2020eep, Liu:2020akv}, and the other types of groups \cite{Kikuchi:2020nxn, Almumin:2021fbk, Ding:2021iqp, Feruglio:2021dte, Kikuchi:2021ogn, Novichkov:2021evw} in which masses, mixing, and CP phases for the quark and/or lepton have been predicted~\footnote{For interested readers, we provide some literature reviews, which are useful to understand the non-Abelian group and its applications to flavor structure~\cite{Altarelli:2010gt, Ishimori:2010au, Ishimori:2012zz, Hernandez:2012ra, King:2013eh, King:2014nza, King:2017guk, Petcov:2017ggy}.}.
Moreover, a systematic approach to understanding the origin of CP transformations has been discussed in Ref.~\cite{Baur:2019iai}, 
and CP/flavor violation in models with modular symmetry was discussed in Refs.~\cite{Kobayashi:2019uyt,Novichkov:2019sqv}, 
and a possible correction from K\"ahler potential was discussed in Ref.~\cite{Chen:2019ewa}. Furthermore,
systematic analysis of the fixed points (stabilizers) has been discussed in Ref.~\cite{deMedeirosVarzielas:2020kji}.
%%%%%  定期分(end) %%%%%%% 

In this paper, we study a Zee-Babu model imposing the modular $A_4$ symmetry, in which we further search for predictions from this symmetry. We perform $\Delta \chi^2$ analysis to get our results and show several predictions for both of the normal and inverted hierarchies.

This paper is organized as follows.
In Sec.~II, we review our model, constructing the valid Yukawa Lagrangian,  Higgs potential,  charged-lepton mass matrix, and neutrino mass matrix. In neutrino sector, we show several relations coming from features of Zee-Babu model and modular $A_4$ symmetry. Then, we perform the $\Delta \chi^2$ analysis to get our results, considering their theoretical origin.
We have conclusions and discussion in Sec.~III.
In Appendix, we show helpful formalisms on the neutrino masses and mixings, and the modular symmetry.

% \begin{widetext}
\begin{center} 
\begin{table}[tbh!]%[tbc]
%\begin{tiny}
\begin{tabular}{|c||c|c|}\hline\hline  
  & \multicolumn{2}{c|}{Leptons}   \\ \hline \hline
& ~$[\bar L_{L_e},\bar L_{L_\mu},\bar L_{L_\tau}]$~& ~$ \ell_R$~       \\\hline\hline 
%%%
$SU(2)_L$ & $\bm{2}$  & $\bm{1}$     \\\hline 
$U(1)_Y$   & $\frac12$ & $-1$       \\\hline
 $A_4$ & $[1,1'',1']$ & $3$         \\ \hline
$-k_I$ & $[-2,-4,-6]$ & $0$    \\
\hline
\end{tabular}
\caption{Lepton contents and  their charge assignments in Zee-Babu model under $SU(2)_L\times U(1)_Y\times A_4$ where $-k_I$ is the number of modular weight and  $\ell_R\equiv [e_R,\mu_R,\tau_R]^T$.}
\label{tab:1}
% \end{tiny}
\end{table}
\end{center}
%\end{widetext}

\section{Model}
\begin{table}[t!]
\begin{tabular}{|c||c|c|c|c|c|}
\hline\hline  
                   & ~$H$~  & ~${s^-}$~  & ~$k^{++}$~ \\\hline 
$SU(2)_L$ & $\bm{2}$   & $\bm{1}$ & $\bm{1}$    \\\hline 
$U(1)_Y$   & $\frac12$  & $1$ & $2$     \\\hline
$A_4$   & $\bm{1}$     & $\bm{1}$  & $\bm{1}$    \\\hline
$-k_I$   & $0$  & $-4$   & $0$     \\\hline
\end{tabular}
\caption{\small 
Charge assignments for boson sector in Zee-Babu model under $SU(2)_L\times U(1)_Y$.}
\label{tab:2}
\end{table}

 \subsection{Model setup}
Here, we review Zee-Babu mode, and assigning charges under the symmetries of $SU(2)_L\times U(1)_Y\times A_4 \times (-k)$ into the lepton and Higgs sectors where $SU(3)_c$ is singlet for any exotic fields.
As for the fermion sector, no new fields are introduced.
%%%
In the SM fermions, the left-handed leptons, which are denoted by $\bar L_L\equiv[\bar L_{L_e},\bar L_{L_\mu},\bar L_{L_\tau}]$, are respectively assigned to be $[1,1'',1']$ under $A_4$ with $[-2,-4,-6]$ modular weight~\footnote{This is one of the minimum assignments. In fact, modular weight combinations $[-2,-2,-2]$, $[-2,-2,-4]$, $[-2,-4,-4]$ for $[\bar L_{L_e},\bar L_{L_\mu},\bar L_{L_\tau}]$ do not satisfy the neutrino oscillation data.}, while right-handed charged-leptons symbolized by $\ell_R\equiv [e_R,\mu_R,\tau_R]^T$ are assigned to be triplet under $A_4$ with zero modular weight.
The fermionic contents and their charged assignments are shown in Table~\ref{tab:1}.

In the boson sector, we introduce a singly-charged scalar $s^-$ with $-4$ modular weight, and
a doubly-charged singlet one $k^{++}$ with zero modular weight, where all the bosons including SM Higgs $H$ are assigned to be trivial singlet $1$ under $A_4$.
$s^-$ and $k^{++}$ play an role in generating the Zee-Babu two-loop neutrino mass matrix.
The SM Higgs is defined by $H=[h^+,(v_H+h_0+i z)/\sqrt2]^T$, where $h^+$ and $z$ are absorbed by the SM gauge bosons $W^+$ and $Z$, respectively and $v_H$ is 246 GeV. The bosonic field contents and their charge assignments are listed in Table~\ref{tab:2}.
Under these symmetries, we find the valid renormalizable Lagrangian as follows:
\begin{align}
  -\mathcal{L}_Y & = 
a_\ell \bar L_{L_e} H (y_1 e_R + y_2 \tau_R + y_3 \mu_R)
+b_\ell \bar L_{L_\mu} H (y_3^{(4)} \tau_R + y_1^{(4)} \mu_R + y_2^{(4)} e_R)\nn\\
&+c_\ell \bar L_{L_\tau} H (y_2^{(6)} \mu_R + y_1^{(6)} \tau_R + y_3^{(6)} e_R)
+d_\ell \bar L_{L_\tau} H (y_2^{'(6)} \mu_R + y_1^{'(6)} \tau_R + y_3^{'(6)} e_R)
 \nn\\
  & +a \bar L_{L_e} (i\sigma_2) L^C_{L_\mu} s^-
  +b \bar L_{L_e} (i\sigma_2) L^C_{L_\tau} s^-
  + c \bar L_{L_\mu} (i\sigma_2) L^C_{L_\tau} s^-
   \nn\\
  & +g (\bar e^C_{R} e_{R} +  \bar \mu^C_{R} \tau_{R} + \bar \tau^C_{R} \mu_{R} ) k^{++}
+  {\rm h.c.} ,
\label{yukawa}
\end{align}
where $Y^{(2)}_3\equiv[y_1,y_2,y_3]^T$, $Y^{(4)}_3\equiv[y_1^{(4)},y_2^{(4)},y_3^{(4)}]^T$, $Y^{(6)}_3\equiv[y_1^{(6)},y_2^{(6)},y_3^{(6)}]^T$, $Y^{'(6)}_3\equiv[y_1^{'(6)},y_2^{'(6)},y_3^{'(6)}]^T$~\footnote{See Appendix.}, 
$a\equiv a' Y^{(10)}_{1'}$, $b \equiv  b' Y^{(12)}_{1''}$, $c \equiv  c' Y^{(14)}_{1}$, 
and $\sigma_2$ is the second component of the Pauli matrix.
Notice here that we have five real parameters; $a_\ell, b_\ell, c_\ell, a, g$ and  the three complex ones;
$d_\ell, b,c$ after phase redefinition of fields without loss of generality in the Yukawa sector.
$a_\ell, b_\ell,c_\ell$ are determined in order to fix the masses of charged-lepton.
To realize the anti-symmetric terms $\bar L_L(i\sigma_2)L^C_L s^-$, Yukawas with modular weight 8 is minimally requested. This is because it appears three independent nonzero singlets $1,1',1''$ as the minimum modular weight.
\footnote{In other words, the neutrino oscillation data cannot be satisfied due to reducing the matrix rank of neutrino mass matrix to be one, if one would not apply these Yukawas within 8 modular weight. }
%In the terms of $a_\chi,b_\chi,c_\chi$ that are complex parameters, we assume to be $a_\chi>>b_\chi, c_\chi$ so that we can explain the electron g-2 without concerning the LFVs.~\footnote{Although terms of $a_\eta,b_\eta,c_\eta$ can contribute to the electron g-2, the actual value is $10^{-15}$ at most. This is because these terms are restricted by LFVs such as $\mu\to e\gamma$ and neutrino oscillations. Thus, we need another terms by introducing $\chi^+$. }
%%%
The non-trivial  terms in Higgs potential is given by 
\begin{align}
  %%%
  {\cal V} &=
\mu     s^- s^- k^{++}  +
   {\cal V}_{2}^{tri} + {\cal V}_4^{tri}   + {\rm h.c.}
  ,\label{Eq:pot}
\end{align}
where $ \mu\equiv \mu_0 Y^{(8)}_1$ is a complex mass scale parameter contributing to the neutrino mass matrix, ${\cal V}_{2}^{tri}$ and ${\cal V}_4^{tri}$ are respectively trivial quadratic and quartic terms of the Higgs potential;
$
{\cal V}_{2}^{tri}= \sum_{\phi=H,s^-,k^{++},} \mu_{\phi}^2|\phi|^2,\quad
{\cal V}_{4}^{tri}= \sum_{\phi'\le\phi}^{\phi=H,s^-,k^{++}} \lambda_{\phi\phi'} |\phi^\dag\phi'|^2.\label{Eq:pot-tri}
$

\subsection{Charged-lepton mass matrix}
After the spontaneous symmetry breaking of $H$,
the charged-lepton mass matrix $(M_e)_{LR}$ is found as follows:
\begin{align}
(M_e)_{LR}  = \frac{v_H}{\sqrt2}
\begin{pmatrix}
|a_\ell| & 0 & 0 \\ 
 0 &  |b_\ell| &0 \\ 
0 & 0 &  |c_\ell| \\ 
\end{pmatrix}
%%%
\begin{pmatrix}
 y_1 & y_3 & y_2 \\ 
 y_2^{(4)} & y_1^{(4)} & y_3^{(4)} \\ 
y_3^{(6)}+\frac{d_\ell}{ |c_\ell|} y_3^{'(6)} & y_2^{(6)}+ \frac{d_\ell}{ |c_\ell|} y_2^{'(6)}  & y_1^{(6)}+ \frac{d_\ell}{ |c_\ell|} y_1^{'(6)} \\ 
\end{pmatrix}
=\frac{v_H}{\sqrt2} \tilde M_e
 \label{massmat}.
\end{align}
Here, $\tilde M_e$ will be useful to discuss the neutrino sector in the next subsection.
Then, the above matrix is diagonalized by bi-unitary mixing matrix as $D_e\equiv{\rm diag}(m_e,m_\mu,m_\tau)=V^\dag_{eL} (M_e)_{LR} V_{eR}$.
In our numerical analysis, we can solve $a_\ell, b_\ell, c_\ell$ by the following relations,
inputting the experimental masses of charged-lepton masses and fixed $\tau$;
\begin{align}
&{\rm Tr}[(M_e)_{LR}(M_e)_{LR}^\dag] = |m_e|^2 + |m_\mu|^2 + |m_\tau|^2,\\
&{\rm Det}[(M_e)_{LR}(M_e)_{LR}^\dag] = |m_e|^2  |m_\mu|^2  |m_\tau|^2,\\
&({\rm Tr}[(M_e)_{LR}(M_e)_{LR}^\dag])^2 -{\rm Tr}[(M_e)_{LR}(M_e)_{LR}^\dag)^2] =2( |m_e|^2  |m_\nu|^2 + |m_\mu|^2  |m_\tau|^2+ |m_e|^2  |m_\tau|^2 ).
\end{align}

%\if0
\subsection{Active neutrino mass matrix}
\label{neut}
We write the valid Lagrangian for the neutrino sector is found as follows:
\begin{align}
-{\cal L}_\nu &= (-\bar\ell_{L_i} f_{ij} \nu^C_{L_j} +\bar\nu_{L_i} f^T_{ij} \ell^C_{L_j}) s^-
  +\bar e_{R_i}^C g_{ij} e_{R_j}+ \mu s^- s^- k^{++} +   {\rm h.c.} ,\label{eq:neut}\\
f&= \begin{pmatrix}
0  &  a & b \\ 
-a & 0 &c \\ 
-b & -c & 0 \\ 
\end{pmatrix}
=c
\begin{pmatrix}
0  &  \epsilon' & \epsilon \\ 
-\epsilon' & 0 &1 \\ 
-\epsilon & -1 & 0 \\ 
\end{pmatrix}
\equiv c \tilde f
,\quad
g= \begin{pmatrix}
1  &  0 &0 \\ 
0& 0 &1 \\ 
0 & 1 & 0 \\ 
\end{pmatrix},
\end{align}
where $\epsilon'\equiv a/c$ and  $\epsilon\equiv b/c$.
Then, the neutrino mass matrix is given at two-loop level as follows:
\begin{align}
  (m_{\nu})_{ij}
 &\simeq \frac{\mu}{64\pi^4} \frac{v_H^2 c^2}{m_s^2}I(r) [\tilde f^T \tilde M_e^* g\tilde M_e^\dag \tilde f]_{ij}
 \equiv  \mu' [\tilde f^T \tilde M_e^* g\tilde M_e^\dag \tilde f]_{ij} ~,
\label{massmatrix2}\\
%%%
&I(r) = -\int_0^1 dx \int_0^{1-x} dy \frac{1-y}{x+(r-1)y+y^2} \ln\left[\frac{y(1-y)}{x+r y}\right],
\end{align}
where $r\equiv m_k^2/m_s^2$, $m_\nu\equiv \mu'\tilde m_\nu $ and we simplified the above form in terms of charged-lepton mass matrix, assuming $\frac{m_{e,\mu,\tau}}{m_{s,k}}<<1$ that is reasonable approximation since singly- and doubly-charged bosons receive stringent constraints on lower masses at Large Hadron Collider~\cite{lhc-bound}.
Thus, the structure of neutrino mass matrix does not depend on the loop function; the neutrino mass matrix is rewritten by charged-lepton mass matrix $M_e$.
%%%
\if0
Here, making the use of an interesting relation $f (1,-\epsilon,\epsilon')^T=0$, one straightforwardly finds
\begin{align}
 m_{\nu} (1,-\epsilon,\epsilon')^T=0.
\end{align}
On the other hand, 
\fi
%%%
$\tilde m_\nu$ is written by measured parametrization $\tilde m_\nu= U_{\nu} \tilde D_\nu U_{\nu}^T$.
Then, the Pontecorvo-Maki-Nakagawa-Sakata unitary matrix $U_{PMNS}$~\cite{Maki:1962mu} is given by $U_{PMNS}= V_{e_L}^\dag U_\nu$
 and $U_\nu,\ \tilde D_\nu$ are respectively given by
\begin{align}
U_{\nu}& =
\begin{pmatrix}
1  &  0 &0 \\ 
0& c_{23} &s_{23} \\ 
0 & -s_{23} & c_{23} \\ 
\end{pmatrix}
\begin{pmatrix}
c_{13}  &  0 & s_{13}e^{-i\delta_{CP}} \\ 
0& 0 & 0 \\ 
-s_{13}e^{i\delta_{CP}} &0 & c_{13} \\ 
\end{pmatrix}
\begin{pmatrix}
c_{12}  &  s_{12} & 0 \\ 
-s_{12} & c_{12} & 0 \\ 
0&0 & 1 \\ 
\end{pmatrix}
\begin{pmatrix}
1  &  0 &0 \\ 
0& e^{i\alpha/2} &0 \\ 
0 & 0 & 1 \\ 
\end{pmatrix}
,
\\
%%%
\tilde D_\nu& = {\rm diag}[\tilde m_1,\tilde m_2,\tilde m_3],
%\begin{pmatrix}\tilde m_1  &  0 &0 \\ 0& \tilde m_{2} &0 \\ 0 & 0 & \tilde m_3 \\ \end{pmatrix},
\quad (\tilde m_{1(3)}=0\ {\rm for\  NH(IH)}),
\end{align}
and $s(c)_{12,23,13}$ are abbreviations of $\sin(\cos)\theta_{12,23,13}$.
Notice here that we have one Majorana phase and the lightest neutrino mass eigenvalue is zero without loss of generality because $m_\nu$ is matrix rank 2.
%%%%%%%%%
\if0
Therefore, we find the following relations (see Appendix A):
\begin{align}
&{({\rm NH})}:\nn\\
&\epsilon = c_{23}\frac{s_{12}}{c_{12}c_{13}} + s_{23} \frac{s_{13}}{c_{13}}e^{i\delta_{CP}},\nn\\
&\epsilon' = s_{23}\frac{s_{12}}{c_{12}c_{13}} - c_{23} \frac{s_{13}}{c_{13}}e^{i\delta_{CP}},\label{eq:eep_nh}\\
&{({\rm IH})}:\nn\\
&\epsilon = - s_{23} \frac{c_{13}}{s_{13}}e^{i\delta_{CP}},\nn\\
&\epsilon' = c_{23} \frac{c_{13}}{s_{13}}e^{i\delta_{CP}}.\label{eq:eep_ih}
%%%
\end{align}
The above relations suggest that structure of the neutrino mass matrix is uniquely determined except overall mass scale $\mu'$, when we input the experimental values $s_{12,23,13},\delta_{CP}$ and theoretical value $\tau$.  
\fi
%%%%%
Then, we show how to determine the two mass squared differences; $\Delta m^2_{\rm atm}$ and $\Delta m^2_{\rm sol}$, which are observables.
 $\Delta m^2_{\rm atm}$ is called the atmospheric mass squared difference, and  $\Delta m^2_{\rm sol}$ is the solar neutrino mass-squared difference.
% that is easy task for Zee-Babu case. 
At first, the values of $\mu'$ is fixed by inputting $\Delta m^2_{\rm atm}$ as follows:
\begin{align}
({\rm NH}):\  \mu'^2= \frac{|\Delta m_{\rm atm}^2|}{\tilde m_3^2},
\quad
({\rm IH}):\  \mu'^2= \frac{|\Delta m_{\rm atm}^2|}{\tilde m_2^2}.
 \end{align}
Then, $\Delta m^2_{\rm sol}$ is derived in terms of $\mu'^2$ as follows:
\begin{align}
& ({\rm NH}):\   \Delta m_{\rm sol}^2= {\mu'^2} {\tilde m_2^2}
=\frac{\tilde m_2^2}{\tilde m_3^2} |\Delta m_{\rm atm}^2|,\\
& ({\rm IH}):\   \Delta m_{\rm sol}^2= {\mu'^2}({\tilde m_2^2-\tilde m_1^2})
=\left(1-\frac{\tilde m_1^2}{\tilde m_2^2}\right) |\Delta m_{\rm atm}^2|.
\end{align}
Considering $ \Delta m_{\rm sol}^2<< \Delta m_{\rm atm}^2$ in addition to the above relations, {\it we predict degenerate neutrino mass eigenvalues for IH and  hierarchal neutrino mass eigenvalues for NH.}
Thus, we estimate sum of neutrino masses $\sum m_i$ as follows:
\begin{align}
& ({\rm NH}):\   \sum m_i \sim \sqrt{\Delta m_{\rm atm}^2}= {\cal O}(0.05)\ {\rm eV},\\
& ({\rm IH}):\   \sum m_i \sim 2 \sqrt{\Delta m_{\rm atm}^2}= {\cal O}(0.1)\ {\rm eV}.\label{eq:mrel}
\end{align}
 
%Here , we will adopt NuFit 5.0~\cite{Esteban:2020cvm} to our $\Delta \chi^2$ analysis. 
%  
The neutrinoless double beta decay is found as 
\begin{align}
& ({\rm NH}):\ \langle m_{ee}\rangle=\mu'|\tilde m_2 s^2_{12} c^2_{13}
e^{i\alpha_{2}}+\tilde m_{3} s^2_{13}e^{-2i\delta_{CP}}|,\\
& ({\rm IH}):\ \langle m_{ee}\rangle=\mu'|\tilde m_1 c^2_{12} c^2_{13}+\tilde m_2 s^2_{12} c^2_{13}
e^{i\alpha_{2}}|,
\end{align}
which may be able to observed by future experiment of KamLAND-Zen~\cite{KamLAND-Zen:2016pfg}. 
Considering the mass relations for NH(IH), we also estimate the neutrinomass double beta decay inputting experimental value of $s_{12,13}$ as follows:
\begin{align}
& ({\rm NH}):\ \langle m_{ee}\rangle \sim |m_{3} s^2_{13}e^{-2i\delta_{CP}}| = \sqrt{\Delta m_{\rm atm}^2} s^2_{13}
={\cal O}(0.01)\ {\rm eV}
,\\
& ({\rm IH}):\ \langle m_{ee}\rangle\sim 
m_2 |c^2_{12} c^2_{13}+ s^2_{12} c^2_{13} e^{i\alpha_{2}}|\sim  
\sqrt{\Delta m_{\rm atm}^2}|c^2_{12} c^2_{13}+ s^2_{12} c^2_{13} e^{i\alpha_{2}}|,
\end{align}
where IH would not determine the sharp value because of Majorana phase.

%%%%%%%%%%%%%%%%%%%%%%%%%%%%%%%%%%%%%%%%%%%%%%%%%%
\subsection{Numerical analysis \label{sec:NA}}
%%%%%%%%%%%%%%%%%%%%%%%%%%%%%%%%%%%%%%%%%%%%%%%%%%
In this subsection, we present our $\Delta \chi^2$ analysis, adopting the neutrino experimental data at 3$\sigma$ interval in NuFit5.0~\cite{Esteban:2018azc} as follows:
\begin{align}
&{\rm NH}: \Delta m^2_{\rm atm}=[2.431, 2.598]\times 10^{-3}\ {\rm eV}^2,\
\Delta m^2_{\rm sol}=[6.82, 8.04]\times 10^{-5}\ {\rm eV}^2,\\
&\sin^2\theta_{13}=[0.02034, 0.02430],\ 
\sin^2\theta_{23}=[0.407, 0.618],\ 
\sin^2\theta_{12}=[0.269, 0.343],\nn\\
%%%
&{\rm IH}: \Delta m^2_{\rm atm}=[2.412, 2.583]\times 10^{-3}\ {\rm eV}^2,\
\Delta m^2_{\rm sol}=[6.82, 8.04]\times 10^{-5}\ {\rm eV}^2,\\
&\sin^2\theta_{13}=[0.02053, 0.02436],\ 
\sin^2\theta_{23}=[0.411, 0.621],\ 
\sin^2\theta_{12}=[0.269, 0.343].\nn
\end{align}
The charged-lepton masses are assumed to be Gaussian distribution.  
Then, our theoretical input parameters are $\tau, \epsilon, \epsilon', d_\ell$ and we work on fundamental region of $\tau$.

%------------------------------ NH ------------------------------------
\begin{figure}[htbp]
  \includegraphics[width=77mm]{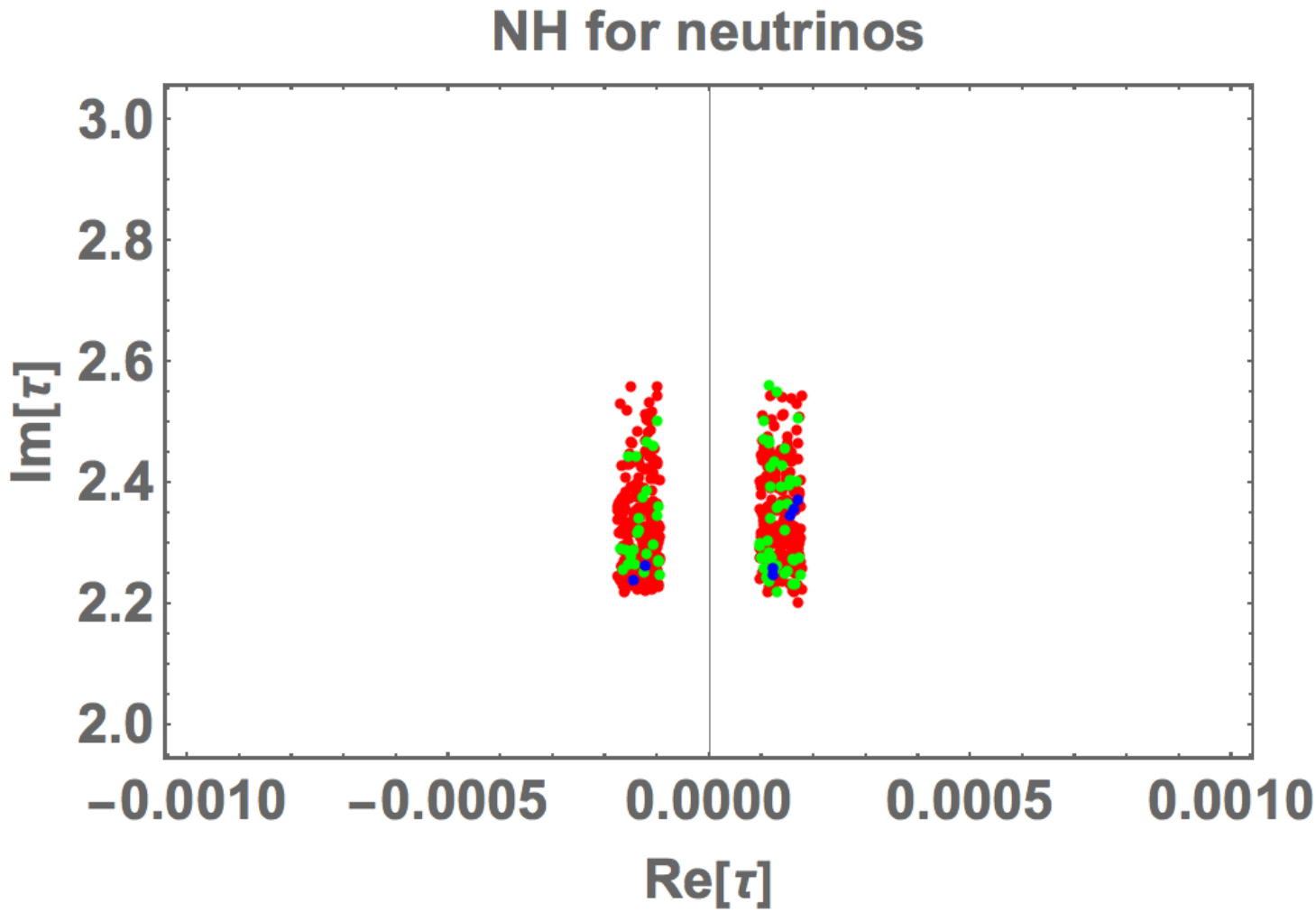}
  \caption{The scatter plots for the real $\tau$ and imaginary $\tau$  in NH.
    In the $\Delta\chi^2$ analysis, the blue color represents the region within 2, green  2-3, and, red 3-5 of $\sqrt{\Delta \chi^2}$.
%    The black solid line is the boundary of fundamental domain at $|\tau|=1$.
}
  \label{fig:tau_nh}
\end{figure}
%%%
\subsubsection{NH}
Figure~\ref{fig:tau_nh} shows the allowed region of $\tau$ within the fundamental region.
It suggests that imaginary part of $\tau$ is localized at nearby [$2.2i$--$2.6i$], while whole the region is allowed for real part of $\tau$, where the blue color represents the region within 2, green  2-3, and, red 3-5 of $\sqrt{\Delta \chi^2}$.
We notice here that real part of $\tau$ tends to be zero, if we focus on the smaller $\Delta \chi^2$.
It is important since $\tau\sim $ [$2.2i$--$2.6i$], is equivalent the region at nearby one of the fixed point $\tau=i\times \infty$~\cite{Okada:2020ukr}.
%In this limit, we get $[y_1,y_2,y_3]^T\sim[1,0,0]^T$.

\begin{figure}[htbp]
  \includegraphics[width=77mm]{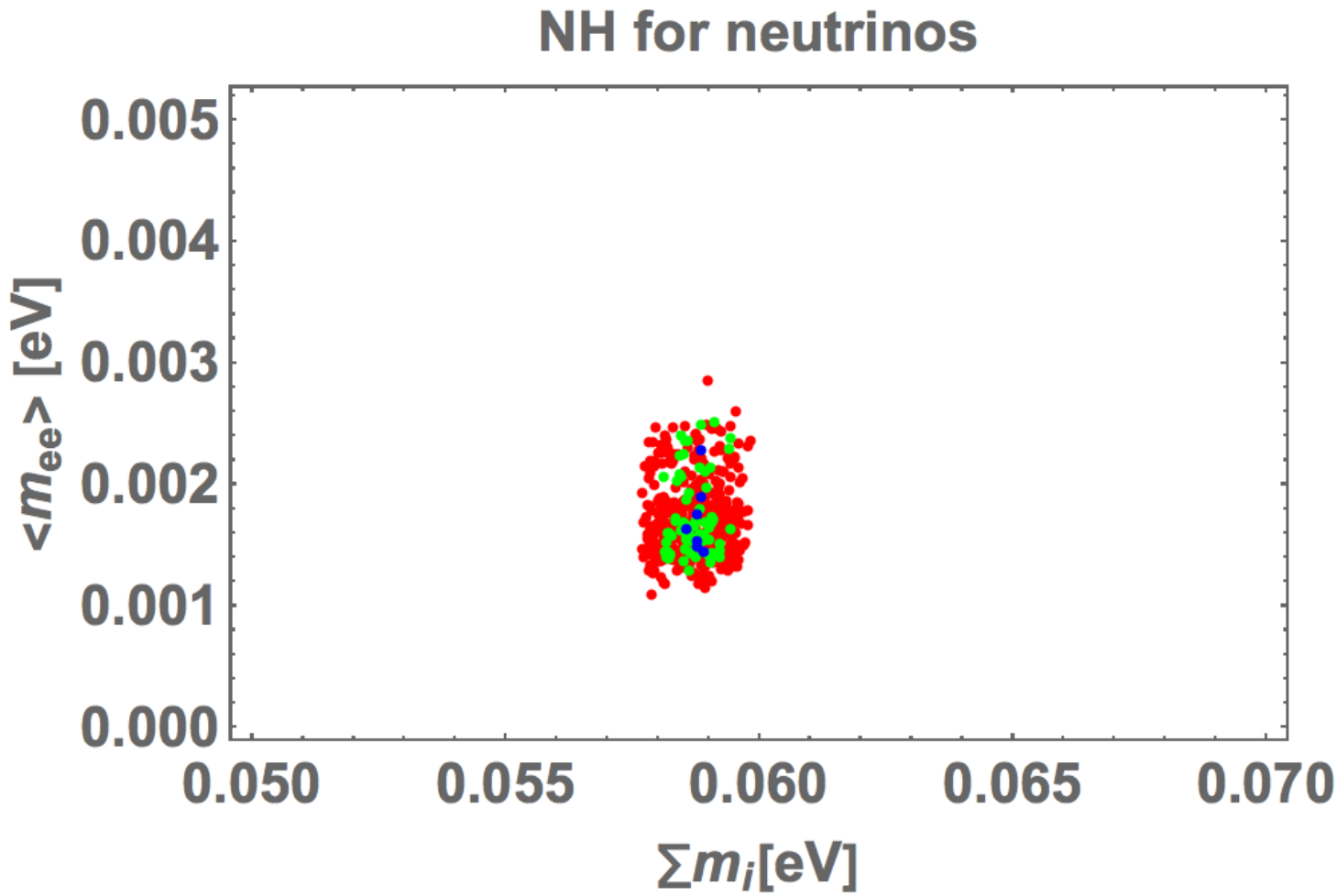}
  \caption{The scatter plots for the neutrinoless double beta decay in terms of sum of neutrino masses.
 The legend of color is the same as Fig.1.}
  \label{fig:masses_nh}
\end{figure}
%%%%%%
Figure~\ref{fig:masses_nh} shows correlation  between $\langle m_{ee}\rangle$ and $\sum m_i$, where the legend of color is the same as Fig.1.
It implies that $\sum m_i$ is allowed at the range of [0.058-0.06] eV, and  $\langle m_{ee}\rangle$ [0.001-0.0025] eV,
which is in favor of our theoretical estimation.

\begin{figure}[htbp]
  \includegraphics[width=77mm]{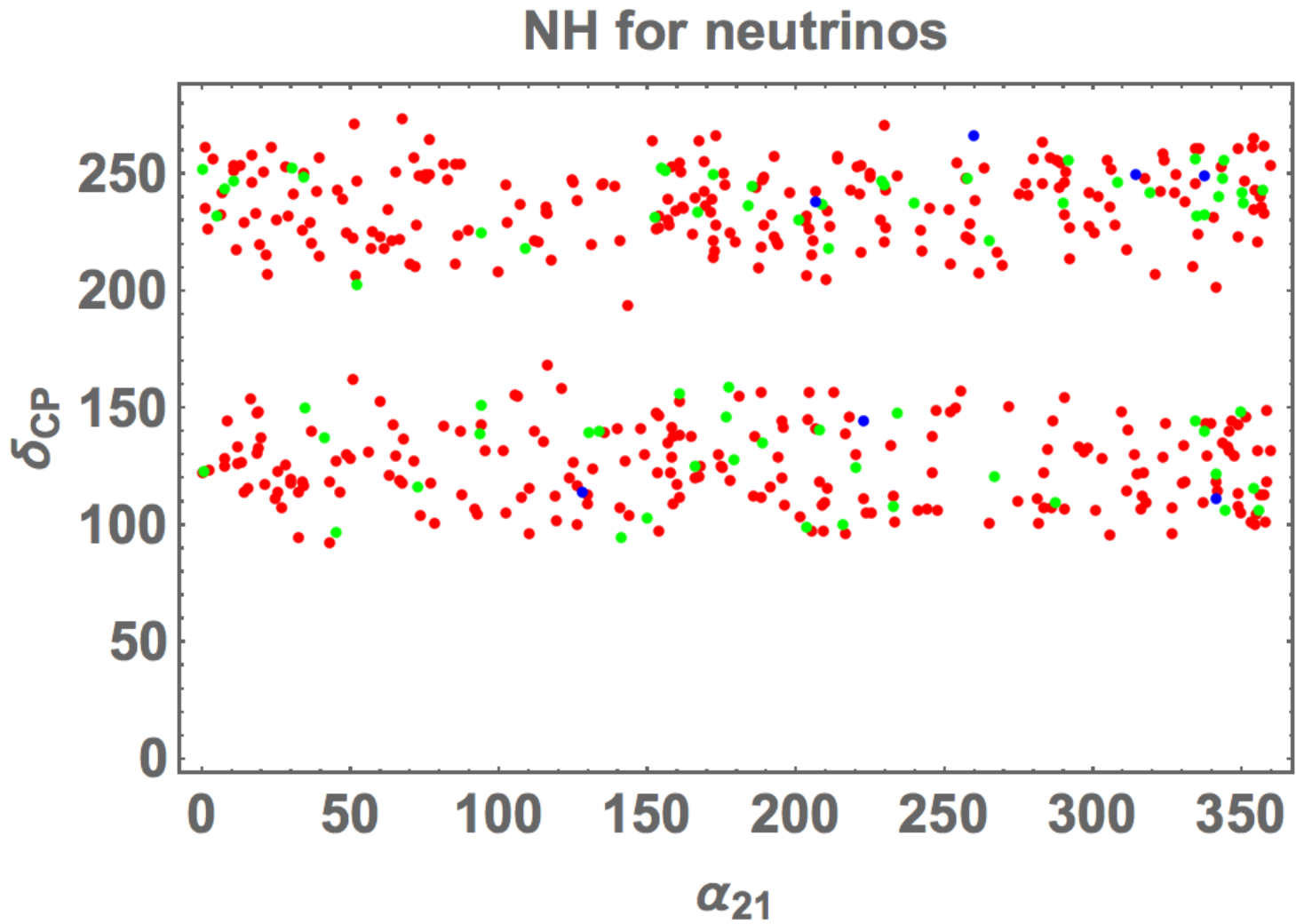}
  \caption{The scatter plots for the Dirac CP phase in terms of Majorana phase.
The legend of color is the same as Fig.1.}
  \label{fig:phase_nh}
\end{figure}
Figure~\ref{fig:phase_nh} shows correlation  between $\delta_{CP}$ and $\alpha_{21}$, where the legend of color is the same as Fig.1.
It reads that Majorana phase tends to be localized at the range of [90-160, 200-270] deg, and whole the range is allowed for  the Dirac CP phase.

\subsubsection{IH}
Since IH needs severe fine-tuning to satisfy the neutrino oscillation data, we show a benchmark point instead of demonstrating figures of NH.

\subsubsection{Benchmark point for NH(IH) and consideration the result}
We also show a benchmark point for NH and IH in Table~\ref{bp-tab}, where we take $\sqrt{\Delta \chi^2}$ is minimum. 
This table suggests that the allowed region of $\tau$ is in favor of the point at nearby $2.3i$ for both the cases.
Here, we study the reason why this region is favored.
In this case, one can approximate $[y_1,y_2,y_3]^T\sim[1,0,0]^T$~\cite{Okada:2020ukr}. 
%%%
Under this limit, we find the mass eigenvalues and vectors for the charged-lepton mass matrix to be $(a_\ell,b_\ell,c_\ell)$ and $V_{e_L}=1_{3\times3}$ respectively,
\footnote{Notice here that there are six more degrees of freedom how to assign $a_\ell,b_\ell, c_\ell$ for the charged-lepton family.
For example, if we assign $a_\ell\sim m_e$, $b_\ell\sim m_\tau$, $c_\ell\sim m_\mu$, $V_{e_L}$ has to be  
$\begin{pmatrix}
1 & 0 & 0 \\ 
0&  0 & 1 \\ 
0 &1 &0 \\ 
\end{pmatrix}$. This degrees of freedom affect $U_\nu$ since $U_\nu=V_{e_L} U_{PMNS}$. }
and parameters $a_\ell,b_\ell$
can be rewritten in terms of the charged-lepton masses and $c_\ell$ as follows:
\begin{align}
a_\ell = \frac{m_e}{m_\tau} c_\ell,\quad
b_\ell = \frac{m_\mu}{m_\tau} c_\ell.
\end{align}
%%%
Inserting the above relations, our neutrino mass matrix simplifies as follows~\footnote{Notice here that $a,b,c$ in $f$ are assumed to be nonvanishing so as to satisfy the neutrino oscillation data even in the limit of $\tau=i\times \infty$.}:
\begin{align}
\tilde m_\nu\sim%b_\ell^2
\begin{pmatrix}
2  \frac{m_\mu}{m_\tau}\epsilon\epsilon'  &   \frac{m_\mu}{m_\tau} \epsilon'  & -  \frac{m_\mu}{m_\tau} \epsilon \\ 
   \frac{m_\mu}{m_\tau} \epsilon'&   \frac{m_e^2}{m_\tau^2} \epsilon'^2 &  - \frac{m_\mu}{m_\tau} +  \frac{m_e^2}{m_\tau^2} \epsilon\epsilon' \\ 
 - \frac{m_\mu}{m_\tau} \epsilon &  - \frac{m_\mu}{m_\tau} +  \frac{m_e^2}{m_\tau^2} \epsilon\epsilon' & \frac{m_e^2}{m_\tau^2} \epsilon^2 \\ 
\end{pmatrix}.
\end{align}
Therefore, we find that
\begin{align}
\frac{(\tilde m_\nu)_{1,2}}{(\tilde m_\nu)_{1,3}}\sim
-\frac{\epsilon'}{\epsilon}.
\end{align}
Substituting the experimental values at 1$\sigma$ into the above equations with the help of Eq.~(\ref{eq:eep_nh}) for NH and Eq.~(\ref{eq:eep_ih}) for IH, we find
\begin{align}
(NH):\left|\frac{\epsilon'}{\epsilon}\right|={\cal O}(0.1-1),
\quad
(IH):\left|\frac{\epsilon'}{\epsilon}\right|={\cal O}(1-10),
\label{eq:theory}
\end{align}
On the other hand, the experimental result for neutrino mass matrix is given by $\tilde m_{\nu}^{exp}\equiv U_{PMNS} \tilde D_\nu U_{PMNS}^T$ as discussed in Sect.~\ref{neut} as follows:
\begin{align}
&({\rm NH}):\quad
\tilde m_{\nu}^{exp}\sim
\begin{pmatrix}
0  &  s_{13}s_{23}e^{i\delta_{CP}} & s_{13} c_{23}e^{i\delta_{CP}} \\ 
s_{13}s_{23}e^{i\delta_{CP}} &  s_{23}^2 &  c_{23} s_{23} \\ 
s_{13}c_{23}e^{i\delta_{CP}} &   c_{23} s_{23} & c_{23}^2 \\ 
\end{pmatrix},\\
%%%
&({\rm IH}):\quad
\tilde m_{\nu}^{exp}\sim
\begin{pmatrix}
c_{12}^2+s_{12}^2 e^{-i\alpha_{21}}  &  s_{12}c_{12}c_{23}(-1+e^{-i\alpha_{21}}) & c_{12}s_{12} s_{23}(1-e^{-i\alpha_{21}}) \\ 
 s_{12}c_{12}c_{23}(-1+e^{-i\alpha_{21}})& * &  * \\ 
 c_{12}s_{12} s_{23}(1-e^{-i\alpha_{21}}) &  * &* \\ 
\end{pmatrix},
\end{align}
where we have approximated them substituting $\tilde m_2<<\tilde m_3$ for NH and $\tilde m_1\sim\tilde m_2$ for IH. 
$*$'s in IH are abbreviated due to complicated forms because they are not important.
Thus, we find 
\begin{align}
&({\rm NH}):\quad
\frac{(\tilde m_\nu^{exp})_{1,2}}{(\tilde m_\nu^{exp})_{1,3}}\sim \frac{s_{23}}{c_{23}},\\
%%%
&({\rm IH}):\quad
\frac{(\tilde m_\nu^{exp})_{1,2}}{(\tilde m_\nu^{exp})_{1,3}}\sim -\frac{s_{23}}{c_{23}}.
\end{align}
Inserting the experimental value for $s_{23}$, one finds 
\begin{align}
\frac{s_{23}}{c_{23}}\sim{\cal O}(1).\label{eq:exp}
\end{align}
From Eq.~(\ref{eq:theory}) and Eq.~(\ref{eq:exp}), one finds that theoretical estimation would overlap with the experimental result.
Hence, we could conclude that allowed region at nearby $\tau=i\times \infty$ is favored for both the cases.

%%%%%%%%%%%%%%%%%%%%%%%%%%%%%%%%%%%%%%%%%%%%%%%%%%%
\begin{table}[h]
	\centering
	\begin{tabular}{|c|c|c|} \hline 
			\rule[14pt]{0pt}{0pt}
%Lepton		&  NH($\tau\approx 1.75 i$) & IH($\tau\approx i$)  & IH($\tau\approx 1.76 i$)\\  \hline
 		&  NH&  IH  \\  \hline
			\rule[14pt]{0pt}{0pt}
		$\tau$&   $ -0.000169 + 2.37 i$ &   $-0.000118+ 4.30 i$   \\ \hline %& $-0.000829+ 1.76 i$\\ 
		\rule[14pt]{0pt}{0pt}
%		$a_\eta$ &$-0.229459 - 1.40234 i$ & $-0.307606 + 0.0129013 I$ & $-4.14934 + 4.25837 I$ \\
		$\frac{\mu'^2}{\rm GeV^2}$ & $1.51\times 10^{-11}$ & $1.66\times 10^{-12}$  \\ \hline
		\rule[14pt]{0pt}{0pt}%
		$[\epsilon,\epsilon']$ & $[-0.0432 - 0.0432 i, 0.248 - 0.0631 i]$ & $[-4.35 - 0.396 i, 4.85 - 0.0573 i]$     \\ \hline
		\rule[14pt]{0pt}{0pt}
%
%		$a$  &  $0.000156626 + 0.000114523 i$ & $3.04135 + 0.98336 I$ & $0.00853417 + 0.0246743 I$\\
		$[a_\ell,\ b_\ell, \ c_\ell]$ & $[3.72\times10^{-6},\ 0.00595,\ 0.000706]$   & $[2.79\times10^{-6},\ 0.000586,\ 0.0100]$    \\ \hline
		\rule[14pt]{0pt}{0pt}
		
%		$a$  &  $0.000156626 + 0.000114523 i$ & $3.04135 + 0.98336 I$ & $0.00853417 + 0.0246743 I$\\
		$d_\ell$ & $65.8 - 117 i$   & $-42.8- 4.36 i$    \\ \hline
		\rule[14pt]{0pt}{0pt}

		$\Delta m^2_{\rm atm}$  &  $2.52\times10^{-3} {\rm eV}^2$   &  $2.50\times10^{-3} {\rm eV}^2$   \\ \hline
		\rule[14pt]{0pt}{0pt}
%		$c'$  &  $-0.213667 - 0.271705 i$ & $-0.153721 + 0.0313641 I$ & $-3.40695 \times 10^{-6} - 0.0000598624 I$\\
		$\Delta m^2_{\rm sol}$  &  $7.30\times10^{-5} {\rm eV}^2$&  $7.43\times10^{-5} {\rm eV}^2$    \\ \hline
		\rule[14pt]{0pt}{0pt}
%		$\sin^2\theta_{12}$ & $ 0.322231$& $0.28036$ & $0.33157$\\
		$\sin\theta_{12}$ & $ 0.556$& $ 0.583$  \\ \hline
		\rule[14pt]{0pt}{0pt}
%		$\sin^2\theta_{23}$ &  $ 0.563489$& $0.463487$ & $0.579596$\\
		$\sin\theta_{23}$ &  $ 0.685$  &  $ 0.666$  \\ \hline
		\rule[14pt]{0pt}{0pt}
%		$\sin^2\theta_{13}$ &  $ 0.0235092$&$0.0240532$ & $0.0218055$\\
		$\sin\theta_{13}$ &  $ 0.148$&  $ 0.151$  \\ \hline
		\rule[14pt]{0pt}{0pt}
%		$\delta_{CP}^\ell$ &  $328.932^\circ$& $ 170.523^\circ$ & $335.678^\circ$\\
		$[\delta_{CP}^\ell,\ \alpha_{21}]$ &  $[238^\circ,\, 207^\circ]$  &  $[178^\circ,\, 179^\circ]$   \\ \hline
		\rule[14pt]{0pt}{0pt}
%		$[\alpha_{21},\,\alpha_{31}]$ &  $[169.96^\circ,\, 336.337^\circ]$ & $[ 167.63^\circ,\, 159.805^\circ]$ &  $[ 157.025^\circ,\, 130.115^\circ]$	\\	
%		$[\alpha_{21},\,\alpha_{31}]$ &  $[0.878^\circ,\, 352^\circ]$   \\	 \hline		\rule[14pt]{0pt}{0pt}
%		$\sum m_i$ &  $71.3811$\,meV &	 $105.101$\,meV & $113.717$\, meV \\
		$\sum m_i$ &  $58.8$\,meV &  $99.3$\,meV     \\ \hline
		\rule[14pt]{0pt}{0pt}
		$\langle m_{ee} \rangle$ &  $1.48$\,meV  &  $15.2$\,meV     \\ \hline
		\rule[14pt]{0pt}{0pt}
		$\sqrt{\Delta\chi^2}$ &  $1.48$ &  $4.99$   \\ \hline
		\hline
	\end{tabular}
	\caption{Numerical benchmark point of our input parameters and observables in NH and IH, where it is taken such that $\sqrt{\Delta \chi^2}$ be minimum.}
	\label{bp-tab}
\end{table}
%%%%%%%%%%%%%%%%%%%%%%%%%%%%%%%%%%%%%%%%%%%%%%%%%%%%%%%
%

\section{Conclusions and discussions}
%%%
We have studied Zee-Babu model with a modular $A_4$ symmetry, and searched for predictions under a minimum framework of the charge assignments.
We have found that $\tau\sim i\times \infty$ is favored to satisfy the neutrino oscillation data for both of the cases for NH and IH.
This is theoretically explainable because it leads to simplified Yukawa couplings $[y_1y_2, y_3]^T=[1,0,0]^T$.
Also, we have a specific patterns for phases for NH and IH.
The sum of neutrino masses and neutrinoless double beta decay are determined by two experiments of  mass squared differences that is direct consequence for Zee-Babu model, which gives rank two neutrino mass matrix.
Here, we have also searched for several non-minimum assignments of modular weight that satisfy the neutrino oscillation data, and have found several models as can be seen in Table~\ref{list-model}, where the other assignments are the same as the minimum one. 
Interestingly, all the models favor the fixed point at near by $\tau=i\times \infty$, even though the other predictions are different each other.

% \begin{widetext}
\begin{center} 
\begin{table}[tbh!]%[tbc]
%\begin{tiny}
\begin{tabular}{|c||c|c||c|c|c|}\hline\hline  
  & \multicolumn{2}{c|}{Leptons}   & \multicolumn{3}{c|}{Bosons}   \\ \hline \hline
& ~$[\bar L_{L_e},\bar L_{L_\mu},\bar L_{L_\tau}]$~& ~$ \ell_R$~& ~$ H$~& ~$ s^-$~& ~$ k^{++}$~       \\\hline\hline 
%%%
% $A_4$ & $[1,1'',1']$ & $3$ & $1$& $1$& $1$         \\ \hline
$-k_I$ & $[-2,-6,-6]$ & $0$& $0$ & $-4$ & $0$    \\ \hline
$-k_I$ & $[-2,-2,-4]$ & $-2$& $0$ & $-4$ & $0$    \\ \hline
$-k_I$ & $[-3,-3,-5]$ & $-1$& $0$ & $-4$ & $0$    \\ \hline
\end{tabular}
\caption{The other models that satisfy the neutrino oscillation data. }
\label{list-model}
% \end{tiny}
\end{table}
\end{center}
%\end{widetext}

%%%
Since modular symmetries are frequently discussed in a supersymmetric theory, we mention the supersymmetric extension of the Zee-Babu model.
In fact, Zee-Babu model can be extended to a supersymmetric theory, introducing three more chiral super fields with opposite $U(1)_Y$ charge of $s^-,k^{++}, H$~\cite{Aoki:2010ib}. In this case, there are two more diagrams contributing to the neutrino mass matrix at two-loop level from  supersymmetric fields. Supposing supersymmetric breaking scale to be much higher than the electroweak scale, we expect supersymmetric contributions are neglected. Thus, we work on the model in a non-supersymmetric theory.
%%%

Before closing our paper, it would be worthwhile discussing the other phenomenologies. 
Typically, Zee-Babu model requires several constraints via lepton universality, lepton flavor violations, and collider physics such as Large Hadron Collider(LHC). As a result, Yukawa couplings $f$ and $g$, and masses of $s^\pm$ and $k^{\pm\pm}$ receive constraints. Detail analyses have already been done by several authors~\cite{Babu:2002uu, AristizabalSierra:2006gb,Schmidt:2014zoa, Nebot:2007bc}.
In our analysis, however, we do not need to consider these constraints , since whole the constraints are 
involved in $\mu'$ that has free parameters. Thus, we can control this value whatever we want.

%%%%%%%%%%%%%%%%%%%%%%%%%%%%%%%%%%%
\vspace{0.5cm}
\hspace{0.2cm} 

\begin{acknowledgments}
The work of H.O. was supported by the Junior Research Group (JRG) Program at the Asia-Pacific Center for Theoretical
Physics (APCTP) through the Science and Technology Promotion Fund and Lottery Fund of the Korean Government and was supported by the Korean Local Governments-Gyeongsangbuk-do Province and Pohang City.
H.O. is sincerely grateful for all the KIAS members.
Y. H. Q. is supported by the Korean Ministry of Education, Science and Technology, Gyeongsangbuk-do Provincial Government, and Pohang City Government for Independent Junior Research Groups at the Asia Pacific Center for Theoretical Physics (APCTP).
\end{acknowledgments}

\section*{Appendix A}
Here, using an interesting relation $f (1,-\epsilon,\epsilon')^T=0$, one straightforwardly finds
\begin{align}
 m_{\nu} (1,-\epsilon,\epsilon')^T=0.\nn
\end{align}
Supposing $V_{e_L}=1$, the neutrino mass matrix $ m_{\nu}= U_{PMNS} [m_1,m_2,m_3] U_{PMNS}^T$ is written by 
\bea
\begin{split}
(m_\nu)_{\text{ee}} &= m_1 c_{12}^2 c_{13}^2 + m_2 c_{13}^2  s_{12}^2+ m_3   s_{13}^2 e^{2 i \delta_{CP} } ,\nn\\
(m_\nu)_{\text{e$\mu $}} & =  m_2 c_{13}  s_{12} (c_{12} c_{23}- s_{12} s_{13} s_{23}e^{-i \delta_{CP} }) - m_1 c_{12} c_{13}  (c_{23} s_{12}+c_{12}  s_{13} s_{23} e^{-i \delta_{CP} } ) + m_3 c_{13}   s_{13} s_{23} e^{i \delta_{CP} } ,\\
(m_\nu)_{\text{e$\tau $}} & =     m_1 c_{12} c_{13}  (s_{12} s_{23}-c_{12} c_{23}  s_{13} e^{-i \delta_{CP} } ) - m_2 c_{13} s_{12} (c_{12} s_{23}+c_{23}  s_{12} s_{13} e^{-i \delta_{CP} } ) + m_3 c_{13} s_{13} c_{23} e^{i \delta_{CP} } , \\
(m_\nu)_{\mu \mu } & =  m_1 (c_{23} s_{12}+c_{12}  s_{13} s_{23} e^{-i \delta_{CP} } )^2+m_2 (c_{12} c_{23}- s_{12} s_{13} s_{23} e^{-i \delta_{CP} })^2+ m_3 c_{13}^2  s_{23}^2, \nn\\
(m_\nu)_{\mu \tau } & = -m_1 (s_{12} s_{23}-c_{12} c_{23} s_{13} e^{-i \delta_{CP} } ) (c_{23} s_{12}+c_{12} s_{13} s_{23} e^{-i \delta_{CP} } ) 
\nn\\
   & -m_2 (c_{12} s_{23}+c_{23} s_{12} s_{13} e^{-i \delta_{CP} } ) (c_{12} c_{23}- s_{12} s_{13} s_{23} e^{-i \delta_{CP} } )+ m_3 c_{13}^2 c_{23} s_{23} , \\
(m_\nu)_{\tau \tau } & =  m_1 (s_{12} s_{23}-c_{12} c_{23} s_{13} e^{-i \delta_{CP} } )^2+ m_2 (c_{12} s_{23}+c_{23}  s_{12} s_{13} e^{-i \delta_{CP} } )^2+ m_3 c_{13}^2 c_{23}^2  .
\end{split}
\eea
Then, the parameters $\epsilon, \epsilon'$ are found to be
\bea
\begin{split}
\epsilon &= \frac{c_{13}  \left[s_{13} s_{23} \left(-2 m_1 m_2+2 m_3  (c_{12}^2 m_2+m_1 s_{12}^2) e^{2 i \delta_{CP}} \right)-2
   c_{12} c_{23} \left(m_1-m_2\right) m_3 s_{12} e^{i \delta_{CP}}\right]e^{i \delta_{CP}} }{2 m_1 m_2 s_{13}^2+2 c_{13}^2 m_3 \left(c_{12}^2 m_2+m_1
   s_{12}^2\right)e^{2 i \delta_{CP}} }, \nn\\
\epsilon' & = \frac{c_{13}  \left[ c_{23} s_{13} \left(m_1 m_2-m_3 (c_{12}^2 m_2+m_1
   s_{12}^2 ) e^{2 i \delta_{CP}} \right)-c_{12} \left(m_1-m_2\right) m_3 s_{12} s_{23} e^{i \delta_{CP}}\right] e^{i \delta_{CP}} }{m_1 m_2 s_{13}^2+c_{13}^2 m_3
   (c_{12}^2 m_2+m_1 s_{12}^2) e^{2 i \delta_{CP}} } .\nn \\
\end{split}
\eea
For NH(IH), i.e., $m_{1(3)}=0$, one has
\begin{align}
&{({\rm NH})}:\nn\\
&\epsilon = c_{23} \frac{ s_{12}}{c_{12} c_{13}}+ s_{23} \frac{s_{13}  }{c_{13}} e^{i \delta _{\text{CP}}} ,
\quad \epsilon'= s_{23} \frac{s_{12} }{c_{12} c_{13}}- c_{23}  \frac{s_{13}  }{c_{13}} e^{i \delta_{\text{CP}}} ,
\label{eq:eep_nh}
\\
&{({\rm IH})}:\nn\\
&\epsilon =  - s_{23} \frac{c_{13} }{s_{13}}  e^{i \delta _{\text{CP}}} , \quad
\epsilon' = c_{23} \frac{c_{13}  }{s_{13}} e^{i \delta _{\text{CP}}} .  \label{eq:eep_ih}
%%%
\end{align}

\section*{Appendix B}

 %%%%%%%%%%%%%%%%%%%%%%%%%%%%%%%%%%%%%%%%%%%%%%%%%%%%%%%%%%%
In this appendix, we present several properties of the modular $A_4$ symmetry. 
In general, the modular group $\bar\Gamma$ is a group of the linear fractional transformation
$\gamma$, acting on the modulus $\tau$ 
which belongs to the upper-half complex plane and transforms as
\begin{equation}\label{eq:tau-SL2Z}
\tau \longrightarrow \gamma\tau= \frac{a\tau + b}{c \tau + d}\ ,~~
{\rm where}~~ a,b,c,d \in \mathbb{Z}~~ {\rm and }~~ ad-bc=1, 
~~ {\rm Im} [\tau]>0 ~.
\end{equation}
This is isomorphic to  $PSL(2,\mathbb{Z})=SL(2,\mathbb{Z})/\{I,-I\}$ transformation.
Then modular transformation is generated by two transformations $S$ and $T$ defined by:
\begin{eqnarray}
S:\tau \longrightarrow -\frac{1}{\tau}\ , \qquad\qquad
T:\tau \longrightarrow \tau + 1\ ,
\end{eqnarray}
and they satisfy the following algebraic relations, 
\begin{equation}
S^2 =\mathbb{I}\ , \qquad (ST)^3 =\mathbb{I}\ .
\end{equation}
More concretely, we fix the basis of $S$ and $T$ as follows:
  \begin{align}
S=\frac13
 \begin{pmatrix}
 -1 & 2 & 2  \\
 2 & -1 & 2  \\
 2 & 2 & -1  \\
 \end{pmatrix} ,\quad 
 T= 
 \begin{pmatrix}
 1 & 0 & 0 \\
0 & \omega & 0  \\
0 & 0 & \omega^2  \\
 \end{pmatrix} ,
 \end{align}
where $\omega\equiv e^{2\pi i/3}$.

Thus, we introduce the series of groups $\Gamma(N)~ (N=1,2,3,\dots)$ that is so-called "principal congruence subgroups of $SL(2,Z)$", which are defined by
 \begin{align}
 \begin{aligned}
 \Gamma(N)= \left \{ 
 \begin{pmatrix}
 a & b  \\
 c & d  
 \end{pmatrix} \in SL(2,\mathbb{Z})~ ,
 ~~
 \begin{pmatrix}
  a & b  \\
 c & d  
 \end{pmatrix} =
  \begin{pmatrix}
  1 & 0  \\
  0 & 1  
  \end{pmatrix} ~~({\rm mod}~N) \right \}
 \end{aligned},
 \end{align}
and we define $\bar\Gamma(2)\equiv \Gamma(2)/\{I,-I\}$ for $N=2$.
Since the element $-I$ does not belong to $\Gamma(N)$
  for $N>2$ case, we have $\bar\Gamma(N)= \Gamma(N)$,
  that are infinite normal subgroup of $\bar \Gamma$ known as principal congruence subgroups.
   We thus obtain finite modular groups as the quotient groups defined by
   $\Gamma_N\equiv \bar \Gamma/\bar \Gamma(N)$.
For these finite groups $\Gamma_N$, $T^N=\mathbb{I}$  is imposed, and
the groups $\Gamma_N$ with $N=2,3,4$ and $5$ are isomorphic to
$S_3$, $A_4$, $S_4$ and $A_5$, respectively \cite{deAdelhartToorop:2011re}.

Modular forms of level $N$ are 
holomorphic functions $f(\tau)$ which are transformed under the action of $\Gamma(N)$ given by
\begin{equation}
f(\gamma\tau)= (c\tau+d)^k f(\tau)~, ~~ \gamma \in \Gamma(N)~ ,
\end{equation}
where $k$ is the so-called as the  modular weight.

%Here we discuss the modular symmetric theory framework without imposing supersymmetry explicitly, considering the $A_4$ ($N=3$) modular group. 
Under the modular transformation in Eq.(\ref{eq:tau-SL2Z}) in case of $A_4$ ($N=3$) modular group, a field $\phi^{(I)}$ is also transformed as 
\begin{equation}
\phi^{(I)} \to (c\tau+d)^{-k_I}\rho^{(I)}(\gamma)\phi^{(I)},
\end{equation}
where  $-k_I$ is the modular weight and $\rho^{(I)}(\gamma)$ denotes a unitary representation matrix of $\gamma\in\Gamma(3)$ ($A_4$ representation).
Thus Lagrangian such as Yukawa terms can be invariant if sum of modular weight from fields and modular form in corresponding term is zero (also invariant under $A_4$ and gauge symmetry).

The kinetic terms and quadratic terms of scalar fields can be written by 
\begin{equation}
 \frac{|\partial_\mu\phi^{(I)}|^2}{(-i\tau+i\bar{\tau})^{k_I}} ~, \quad  \frac{|\phi^{(I)}|^2}{(-i\tau+i\bar{\tau})^{k_I}} ~,
\label{kinetic}
\end{equation}
which is invariant under the modular transformation and overall factor is eventually absorbed by a field redefinition consistently.
Therefore the Lagrangian associated with these terms should be invariant under the modular symmetry.

The basis of modular forms with weight 2, $Y^{(2)}_3 = (y_{1},y_{2},y_{3})$, transforming
as a triplet of $A_4$ is written in terms of Dedekind eta-function  $\eta(\tau)$ and its derivative \cite{Feruglio:2017spp}:
%%%%%%%%%%%%%%%%%%%%%%%
\begin{align} 
\label{eq:Y-A4}
y_{1}(\tau) &= \frac{i}{2\pi}\left( \frac{\eta'(\tau/3)}{\eta(\tau/3)}  +\frac{\eta'((\tau +1)/3)}{\eta((\tau+1)/3)}  
+\frac{\eta'((\tau +2)/3)}{\eta((\tau+2)/3)} - \frac{27\eta'(3\tau)}{\eta(3\tau)}  \right)\nn\\ 
&\simeq
1+12 q+36 q^2+12 q^3+\cdots,\\
y_{2}(\tau) &= \frac{-i}{\pi}\left( \frac{\eta'(\tau/3)}{\eta(\tau/3)}  +\omega^2\frac{\eta'((\tau +1)/3)}{\eta((\tau+1)/3)}  
+\omega \frac{\eta'((\tau +2)/3)}{\eta((\tau+2)/3)}  \right) , \label{eq:Yi} \nn\\ 
&\simeq
-6q^{1/3} (1+7 q+8 q^2+\cdots),\\
y_{3}(\tau) &= \frac{-i}{\pi}\left( \frac{\eta'(\tau/3)}{\eta(\tau/3)}  +\omega\frac{\eta'((\tau +1)/3)}{\eta((\tau+1)/3)}  
+\omega^2 \frac{\eta'((\tau +2)/3)}{\eta((\tau+2)/3)}  \right)\nn\\ 
&\simeq
-18q^{2/3} (1+2 q+5 q^2+\cdots),
\end{align}
where $q=e^{2\pi i \tau}$, and expansion form in terms of $q$ would sometimes be useful to have numerical analysis.
%%%%%%%%%%%%%%%%%%%%%
% Notice here that any singlet couplings under $A_4$ start from $-k=4$ constructed from the modular forms with $-k=2$ while it is absent if $-k=2$.

Then, we can construct the higher order of couplings under $A_4$ singlets $Y^{(4)}_3,\ Y^{(6)}_3,\ Y^{'(6)}_3$, $Y^{(10)}_{1'}$, $Y^{(12)}_{1''}$, $Y^{(14)}_{1}$:
\begin{align}
& Y^{(4)}_1 = y^2_1+2 y_2 y_3, \ Y^{(4)}_{1'} = y^2_3+2 y_1 y_2, \
 Y^{(6)}_1 = y^3_1+y^3_2 + y^3_3 -3 y_1y_2y_3, \\
 & Y^{(8)}_1 = (y_1^2+ 2 y_2 y_3)^2, \
 Y^{(8)}_{1'} = (y_1^2+ 2 y_2 y_3)(y_3^2+ 2 y_1 y_2), \
 Y^{(8)}_{1''} =(y_3^2+ 2 y_1 y_2)^2,\\
& Y^{(10)}_{1'} = Y^{(4)}_{1'} Y^{(6)}_1,\
Y^{(12)}_{1''} = Y^{(4)}_{1} Y^{(8)}_{1''},\
Y^{(14)}_{1''} = Y^{(6)}_{1} Y^{(8)}_{1},\
\\
 %%%
 Y^{(4)}_3& = ( y^2_1-  y_2 y_3,y^2_3-  y_1 y_2, y^2_2-  y_1 y_3),\\
%%%
Y^{(6)}_3& = ( y^3_1+2y_1 y_2 y_3, y_1^2y_2+2 y^2_2 y_3, y^2_1 y_3+2 y^2_3 y_2),\\
Y^{(6)}_{3'} & = ( y^3_3+2y_1 y_2 y_3, y^2_3 y_1+2 y^2_1 y_2, y^2_3 y_2+2 y^2_2 y_1),
%%%
\end{align}
%%%
where  the above relations are constructed by the multiplication rules under $A_4$ as shown below:
\begin{align}
\begin{pmatrix}
a_1\\
a_2\\
a_3
\end{pmatrix}_{\bf 3}
\otimes 
\begin{pmatrix}
b_1\\
b_2\\
b_3
\end{pmatrix}_{\bf 3'}
&=\left (a_1b_1+a_2b_3+a_3b_2\right )_{\bf 1} 
\oplus \left (a_3b_3+a_1b_2+a_2b_1\right )_{{\bf 1}'} \nonumber \\
& \oplus \left (a_2b_2+a_1b_3+a_3b_1\right )_{{\bf 1}''} \nonumber \\
&\oplus \frac13
\begin{pmatrix}
2a_1b_1-a_2b_3-a_3b_2 \\
2a_3b_3-a_1b_2-a_2b_1 \\
2a_2b_2-a_1b_3-a_3b_1
\end{pmatrix}_{{\bf 3}}
\oplus \frac12
\begin{pmatrix}
a_2b_3-a_3b_2 \\
a_1b_2-a_2b_1 \\
a_3b_1-a_1b_3
\end{pmatrix}_{{\bf 3'}\  } \ , \nonumber \\
%%%
a_{1'}\otimes 
\begin{pmatrix}
b_1\\
b_2\\
b_3
\end{pmatrix}_{\bf 3}
&=a\begin{pmatrix}
b_3\\
b_1\\
b_2
\end{pmatrix}_{\bf 3},\quad
a_{1''}\otimes 
\begin{pmatrix}
b_1\\
b_2\\
b_3
\end{pmatrix}_{\bf 3}
=a\begin{pmatrix}
b_2\\
b_3\\
b_1
\end{pmatrix}_{\bf 3},\nn\\
{\bf 1} \otimes {\bf 1} = {\bf 1} \ , \quad &
{\bf 1'} \otimes {\bf 1'} = {\bf 1''} \ , \quad
{\bf 1''} \otimes {\bf 1''} = {\bf 1'} \ , \quad
{\bf 1'} \otimes {\bf 1''} = {\bf 1} \ .
\end{align}

\if0
Finally, we show the features of fixed points of $\tau=i,\omega, i\times \infty$.\\
\begin{itemize}
\item In case of $\tau=i$, it is invariant under the transformation of $\tau\to -1/\tau$ that corresponds to $S$ transformation.
It implies that there is a remnant $Z_2$ symmetry and its element is given by $\{1,S\}$.
Then, the concrete value of $Y^{(2)}_3$ can be written down by~\cite{Okada:2020ukr}
\begin{align}
Y^{(2)}_3&\simeq1.0025(1,1-\sqrt3,-2+\sqrt3).
\end{align}
 \item In case of $\tau=\omega$, it is invariant under the transformation of $\tau\to -1/(1+\tau)$ that corresponds to $ST$ transformation.
It implies that there is a remnant $Z_3$ symmetry and its element is given by $\{1,ST,(ST)^2\}$.
Then, the concrete value of $Y^{(2)}_3$ can be written down by~\cite{Okada:2020ukr}
\begin{align}
Y^{(2)}_3&\simeq 0.9486(1,\omega, -\frac12\omega^2).
\end{align}
\item In case of $\tau=i\times\infty$, this corresponds to $T$ transformation.
It suggests that there is a remnant $Z_3$ symmetry and its element is given by $\{1,T,T^2\}$.
Then, the concrete value of $Y^{(2)}_3$ can be written down by~\cite{Okada:2020ukr}
\begin{align}
Y^{(2)}_3&\simeq (1,0,0).
\end{align}
\end{itemize}
\fi

%\fi

\end{document}